\documentclass[12pt,letterpaper]{article}
\usepackage{jheppub}
\usepackage{bm}

\title{Firewalls From Double Purity\footnote{This article is based on a seminar given at the CERN Workshop on Black Hole Horizons and Quantum Information, in March 2013. Video is available at \url{http://cds.cern.ch/record/1532382}.}}

\author[a,b]{Raphael Bousso}
\affiliation[a]{Center for Theoretical Physics and Department of Physics,\\
 University of California, Berkeley, CA 94720, U.S.A.}
\affiliation[b]{Lawrence Berkeley National Laboratory, Berkeley, CA 94720,
  U.S.A.}

\abstract{The firewall paradox is often presented as arising from double entanglement, but I argue that more generally the paradox is double purity. Near-horizon modes are purified by the interior, in the infalling vacuum. Hence they cannot also be pure alone, or in combination with any third system, as demanded by unitarity. This conflict arises independently of the Page time, for entangled and for pure states. It implies that identifications of Hilbert spaces cannot resolve the paradox. 

Traditional complementarity requires the unitary identification of infalling matter with a scrambled subsystem of the Hawking radiation. Extending this map to the infalling vacuum overdetermines the out-state. More general complementarity maps  (``$A=R_B$'', ``ER $=$ EPR'') founder when the near-horizon zone is pure. I argue that pure-zone states span the microcanonical ensemble, and that this suffices to make the horizon a special place. 

I advocate that the ability to detect the horizon locally, rather than the degree or probability of violence, is what makes firewalls problematic. Conversely, if the production of matter at the horizon can be dynamically understood and shown to be consistent, then firewalls do not constitute a violation of the equivalence principle.}

\begin{document}
\maketitle

\newpage
\paragraph{Notation} Most variables will be defined in the text. Here is a list of key definitions:\\

\begin{tabular}{ l l }
$b$ & a minable mode in the near-horizon region (the ``zone''); \\
\mbox{} & also, the associated annihilation operator\\
$\tilde b$ & its purification inside the black hole, in the infalling vacuum\\
$\mathbf{b}$ & the Hilbert space of $b$\\
$B(t)$ & the collection of all modes minable at the time $t$\\
$XY$ & a bipartite system\\
$\mathbf{X}\otimes \mathbf{Y}$ & its Hilbert space \\
$S_X$ & the von Neumann entropy of $X$: $S_X= -\mathrm{tr}\, \rho_X\log\rho_X$\\
$F-b$ & the complement of the subsystem $b$ in $F$\\
$\mathbf{F}\oslash \mathbf{b}$ & its Hilbert space
\end{tabular}

\section{Introduction}

Unitarity and the equivalence principle---the central principles of quantum mechanics and of general relativity---come into sharp conflict at the horizon of a black hole. Classically, a black hole formed by collapse quickly evolves to the vacuum Kerr solution. In particular, the neighborhood of the horizon is in the vacuum; intuitively, this is because any matter would rapidly fall into the black hole. The vacuum state at the horizon implies that the Hawking radiation is in a mixed state~\cite{Haw76a}: information is lost.

From the viewpoint of quantum mechanics, the Hawking radiation is the out-state of an S-matrix computed by a path integral. If the in-state was pure, then unitarity demands that the out-state is pure. The validity of this viewpoint is closely related to the consistency of black hole thermodynamics: the Bekenstein-Hawking entropy~\cite{Bek72} allows a generalized second law to operate. (The apparent validity of universal entropy bounds~\cite{Bek81,Cas08,Tho93,Sus95,CEB1} suggests that the generalized second law does indeed hold~\cite{FMW}.) But if no entropy can be lost into a black hole, then it would be surprising if information could be lost. Finally, the AdS/CFT correspondence~\cite{Mal97} reduces the computation of the gravitational S-matrix to manifestly unitary evolution in well-defined quantum theory. Thus, the evidence for unitarity is strong.

Black hole complementarity~\cite{Pre92,SusTho93,SteTho94} was an attempt to reconcile the infalling vacuum with unitarity, by exploiting the fact that certain spacelike related operators in the interior and exterior cannot be accessed by any single observer, allowing their identification. But this appears to fail~\cite{AMPS}: in the theory of the infalling observer alone, unitarity and the validity of effective field theory outside the horizon imply that the horizon cannot be in the vacuum state.\footnote{See~\cite{AMPSS} for an extensive list of subsequent work. Precursors include~\cite{Sor97,Bra09V1,Mat09,Gid11,Gid12,GidShi12}. Ref.~\cite{Sus13} offers a clear exposition of the relevant concepts from quantum information theory.}

\paragraph{Firewalls from Double Entanglement} To demonstrate this problem, Almheiri, Marolf, Polchinski, and Sully (AMPS) considered ``old'' black holes that had lost more than half of their original area. Assuming unitarity, the ``late'' radiation that these black holes will decay into is generically highly entangled with the ``early'' radiation that has already been emitted. In particular, modes in the near-horizon zone of the old black hole are entangled with the early radiation, since they can be mined and thus form a subsystem of the late radiation. The zone consists of modes that are far enough from the horizon to be under semiclassical control, but closer than a Schwarzschild radius. In the infalling vacuum, these modes would be entangled with modes inside the black hole. But this would contradict the monogamy of the entanglement, a general property of quantum mechanics. Hence, they cannot be in the vacuum state. Mode by mode, this implies an energy density controlled by the fundamental cutoff, at the horizon. This is the firewall.

AMPS's elegant exploitation of double entanglement has become deeply embedded in the literature, perhaps to the point of obscuring the generality of the conflict. In fact, the zone need not be entangled with the early radiation or with anything else. Suppose instead that the out-state factorizes into a product state of zone and other degrees of freedom. Then the firewall is even more obvious: since the zone is in a pure state all by itself, it cannot also form a pure entangled state with the interior. In fact, no exact factorization is needed: if the entanglement of the zone with the early Hawking radiation is less than thermal, then the state of the zone by itself is incompatible with its being a subsystem of the infalling vacuum. 

The focus on double entanglement has encouraged an optimistic view of the role that complementarity can play in eliminating firewalls. Suppose that not only the infalling matter, but also the vacuous interior regions are identified with scrambled subspaces of the Hawking radiation. For highly entangled states of the zone with the ``early'' radiation, this would appear to circumvent the AMPS argument. After all, the vacuum, too, is a highly entangled state. A suitable choice of map should allow the reconstruction of the vacuum at the horizon. This strategy is variously called ``$A=R_B$'', or ``ER $=$ EPR''; or in the notation of the present paper, $\tilde B(t)=\hat B(t)$.  Recent proposals include~\cite{VerVer12,PapRaj12,Sus13,NomVar13,VerVer13a,VerVer13b,MalSus13}. In all cases, $\hat B$, the exterior purification of the zone modes $B$, is identified with $\tilde B$, the interior partner modes of $B$ in the infalling vacuum. Thus, the inconsistent double entanglement of $B$ is reduced to a consistent, single entanglement. 

However, none of the above arguments apply if the state of the zone and other exterior systems factorizes. More generally they do not apply if the entropy of the zone is small compared to the thermal entropy. But a complete basis of the microcanonical ensemble of a black hole can be constructed from such states, each of which trivially has a firewall. This alone creates a conflict with the equivalence principle.

\paragraph{Outline and Summary} The purpose of this paper is twofold: first, to examine what complementarity can and cannot achieve; and second, to argue that resolutions that exploit entanglement necessarily fall short, because firewalls arise independently of the degree of entanglement between the near horizon zone and other exterior systems. The arguments presented here will make no reference to such entanglement. Hence they apply equally to young and to old black holes, and equally to black holes in entangled and in pure states.\footnote{This article largely follows~\cite{Bou13}. Since then, interesting papers have appeared which have some overlap and some differences. Ref.~\cite{MarPol13} considers unentangled states in the powerful context of AdS/CFT; Ref.~\cite{Cho13} examines the D1-D5 system to argue that a microcanonical ensemble exists for black holes with positive specific heat. The arguments presented here do not rely on gauge gravity duality and do not restrict to near-extremal or large AdS black holes. Ref.~\cite{AMPSS} also speculates that firewalls are continuously produced at the horizon (though somewhat inside).}

In light of the firewall paradox, it is important to reconsider the need for complementarity, to identify its role, and to understand its limitations. It is instructive to begin with the naive viewpoint that the interior and exterior have independent Hilbert spaces. In Sec.~\ref{sec-dp}, I show that firewalls arise from a conflict between the entangled purity of the vacuum with the purity of the out-state, regardless of the amount of entanglement between any of the exterior subsystems. In the remainder of the paper, I argue that complementarity cannot mitigate this basic conflict sufficiently to reconcile unitarity with the equivalence principle. 

Some form of complementarity is clearly required by unitarity, with or without firewalls. A collapsing star cannot hit a firewall at the event horizon: by causality, a firewall can form only later. Inside the black hole, the star carries the same information as the outgoing Hawking radiation at spacelike separation, in apparent violation of the no-cloning theorem~\cite{WooZur82}. However, no observer can see both copies~\cite{SusTho93b}. Strictly, this does not mean that one has to identify the two Hilbert spaces. (One could merely note that in any single observer's description, only one copy appears.) But it is consistent to do so.

The simplest implementation of complementarity is a unitary map of the Hilbert space of any infalling matter to a (possibly scrambled) subsystem of the Hawking radiation. This is linear, and it is causal, assuming that the information only appears in the radiation once it is too late to reunite it with the infalling matter~\cite{HayPre07}. In Sec.~\ref{sec-simple}, I note that if a unitary complementarity map is applied to the infalling vacuum instead of the infalling matter, then it becomes inconsistent with unitarity of the S-matrix, because the vacuum is a unique state. Hence, a unitary identification of the interior vacuum regions with the Hawking radiation does not resolve the conflict between unitarity of the S-matrix and the equivalence principle.

This has motivated ``stronger'' versions of complementarity, in which the complementarity map is not required to be unitary. Instead, it is allowed to depend on the out-state, so that the infalling vacuum is recovered independently of the out-state. This could lead to problems with causality, since in many situations the image of the infalling vacuum would have to be present outside prior to infall. (The outside copy is hard to access computationally in Haar-typical states~\cite{HarHay13}, but the relevance of this obstruction remains controversial~\cite{Bou13,AMPSS}.)

In Sec.~\ref{sec-strong}, I will focus on a different problem: no form of complementarity can restore the infalling vacuum for states in which the zone is in a pure state. I argue that the statistical interpretation of black hole thermodynamics guarantees that such states form a complete basis for the microcanonical ensemble. I then argue that this is sufficient to establish a violation of the equivalence principle, as the presence of a firewall in a complete basis differs in several respects from acceptable particle detections in curved space. The horizon is a special place.

\paragraph{Two Conclusions} Firewalls appear to violate the equivalence principle, in its formulation as the following statement: {\em The vacuum, on scales smaller than the curvature scale, has the same properties everywhere.} Violating the equivalence principle is a serious problem. But it is, in my view, the {\em only\/} problem. I emphasize this because it has two important implications, which appear not to be universally accepted:
\begin{enumerate} 
\item There is no point in pursuing approaches which merely seek to make the horizon less ``violent'', or to make infall possible in some way for most observers, but in which it still is possible for a local observer to notice the event horizon. There is no principle of nonviolence in Nature.  But if the horizon is a special place---if crossing it can be locally detected with sufficient probability---then no matter how harmless the crossing, the equivalence principle is lost, and with it the foundation of general relativity. 

This criterion is quite selective. It excludes any model that fails to address firewalls in product states: as I argue in Sec.~\ref{sec-strong}, the presence of firewalls in these states alone is already incompatible with the properties of the adiabatic vacuum.

Another example, which will not be discussed in the main portion of the paper, are nonlocal modifications of effective field theory. They must selectively involve the short distance modes just inside and outside the horizon, which form the most violent part of the firewall. But if field theory is mainly modified very close to the horizon, then the horizon becomes a special place because detectors behave differently there.\footnote{For this reason, the validity of effective field theory could be eliminated from the assumptions made by AMPS, if the needlessly weak ``absence of drama'' assumption is replaced by the stronger but essential requirement that the horizon must not be a special place.} (In~\cite{Gid12} this problem manifests itself as a failure of the vacuum to produce the correct Unruh effect. An observer hovering near the horizon will detect considerable excess above the thermal flux~\cite{Unr76} that the same detector would see in any other low-curvature region, at the same acceleration.) 

\item If the horizon is not in the vacuum, then firewalls are perfectly acceptable. The equivalence principle is not violated when I bump into a wall: there is matter there, which makes it a special place. This is obvious; the real problem is to understand how deviations from the vacuum can survive near the horizon of a black hole. Why don't they just fall in? What must be happening is that the firewall is continuously produced, from transplanckian modes near the horizon that are getting stretched into observable size by the exponential redshift. Unlike what we usually assume, those modes apparently do not emerge in the vacuum state. 

In cosmology, new semiclassical modes enter in two ways. As we get older, our past light-cone encompasses ever new regions. Their state is determined by initial conditions, which evidently constrain ultraviolet modes to be in the vacuum. And as the universe expands, regions that were already in our past light-cone increase in volume. Conservation of the stress tensor ensures that stretched unexcited modes remain in the vacuum. 

However, near a Killing horizon after the scrambling time $R\log R$, neither constraint applies. Our past light-cone has disconnected boundary components near every black hole, but no new information enters from the distant past as the area of the component becomes nearly independent of time. Meanwhile, the energy cost of any local excitation at the boundary is arbitrarily redshifted. Apparently the fields exploit this freedom to emerge in a nonvacuum state determined by the most recent infalling matter or infalling zone modes. It will be important to understand how this process is consistent with Lorentz-invariance, and how it reproduces the coarse-grained features of black hole thermodynamics.\footnote{It will also be interesting to investigate its role in cosmology, particularly in the context of the measure problem, where a firewall might explain otherwise puzzling features~\cite{BouFre10c}.}

\end{enumerate} 

To summarize, a fundamental principle cannot break down on occasion, without being completely undermined. But the case for firewalls is sound. Hence, what must break down is not the equivalence principle but the adiabatic vacuum. This would allow substantial new physics in particular settings, such as finite Killing horizons, while preserving general relativity.

\section{Firewalls from Double Purity}
\label{sec-dp}

In this section I give a general argument in favor of firewalls that does not assume maximal entanglement of subsystems of the Hawking radiation. Hence it does not depend on whether the black hole is young or old. 

I will exploit that modes outside the black hole must form an entangled pure state with interior degrees of freedom, if an infalling observer is to find a vacuum at the horizon. This contradicts the purity of the S-matrix out-state, independently of whether the mode in question is pure or entangled with other portions of the Hawking radiation.

I will not appeal to complementarity in this section. But in the following sections I will examine various forms complementarity, in light of the present formulation of the paradox; and I will find that firewalls are still required for unitarity. Hedging the firewall argument against objections that ultimately fall short tends to obscure the origin of the problem. Therefore, it is instructive to begin with the most straightforward setting, with all systems treated as distinct.
  
\subsection{Quantum Mechanics Argument}
\label{sec-qmdp}

The argument is general at the level of quantum mechanics, and I will state it abstractly before applying it to the black hole. Consider a bipartite system $XY$ in a pure, entangled (but not necessarily maximally entangled) state:
\begin{equation}
S_{XY}=0~,~~~ S_Y>0~.
\label{eq-sabb}
\end{equation}
Then there cannot exist a third system $Z$, entirely distinct from $X$, such that the state of $YZ$ is pure; for otherwise we would have 
\begin{equation}
S_{YZ}=0~,
\label{eq-sbc}
\end{equation}
and the strong subadditivity of the entanglement entropy~\cite{LieRus73},
\begin{equation}
S_{XYZ}+S_Y\leq S_{XY}+S_{YZ}~,
\label{eq-ss}
\end{equation}
would be violated. 

Conversely, it follows for any $Y$, $Z$: if the state of $YZ$ is pure (entangled or not), then Eq.~(\ref{eq-sabb}) cannot hold, i.e., $Y$ cannot form an entangled (maximally or not) pure state with some other system $X$.\footnote{If $Y$ were {\em maximally\/} entangled with $Z$ then the stronger conclusion would obtain that $Y$ cannot even be classically correlated with $X$. But this conclusion is nowhere needed in the firewall argument.} 

Below, the role of $XY$ will be played by the infalling vacuum; the role of $YZ$ by the final out-state (the entire Hawking radiation). The inference 
\begin{equation}
S_{XY}=0\, \land\, S_Y>0 \Rightarrow S_{YZ}\neq 0,~ \forall Z: X\cap Z=\varnothing
\end{equation}
states that the infalling vacuum implies information loss. The equivalent inference
\begin{equation}
S_{YZ}=0 \Rightarrow \lnot \exists X: (X\cap Z=\varnothing\land S_{XY}=0\,\land\,S_Y>0)
\end{equation}
states that unitarity implies a firewall.

\subsection{Application to Black Holes}
\label{sec-bhdp}

I will now explain how the abstract subsystems and assumptions above are related to physical systems and conditions in a black hole spacetime. An important physical ingredient that was necessarily absent in the abstract argument above is {\em minability\/} of a large class of modes near the horizon. These are the modes that form the firewall. I will pay special attention to the role of mining.

Consider a black hole of radius $R\gg 1$,\footnote{Planck units are used throughout.} formed from a pure state $|\Psi \rangle$. Let $b$ be a mode with support strictly outside the horizon, at a time much greater than $R \log R$ after the most recent infall of matter into the black hole. To be physical, we can work with wave-packets, which can be constructed from the standard exterior mode set.  The modes of greatest relevance for the firewall argument have Killing frequency of order the Hawking temperature, $R^{-1}$; they are localized in the near horizon zone (within $r<3M$), at a proper distance from the horizon not much greater than their characteristic proper size. All scales are assumed much greater than the Planck length, so that effective field theory should describe the mode to good approximation.  

\subsubsection{The Minable Zone as a Subsystem of the Final State}

Modes of this type can be mined~\cite{UnrWal82}: if they are extracted from the zone, the mass of the black hole decreases and the energy of matter far from the black hole increases. Hence, the mode $b$ must be considered a subsystem of the final state, independently of whether it is actually probed:
\begin{equation}
\mathbf{F} = \mathbf{b} \otimes (\mathbf{F} \oslash \mathbf{b})
\label{eq-outboutb}
\end{equation}
Here, $\mathbf{F}$ denotes the Hilbert space for the out-state of the S-matrix, and $\mathbf{F} \oslash \mathbf{b}$ denotes the Hilbert space of the system that complements $\mathbf{b}$ in the final state. The Hilbert space $\mathbf{b}$ is spanned, e.g., by the eigenstates of $b$-occupation number, $|n\rangle_b$. 

Because the mode $b$ can be mined, it is irrelevant whether it is outgoing or not, nor does it matter whether it is an s-wave or has angular momentum.  I assume only that effective field theory is valid outside the horizon, so that $b$ (and, if necessary, the mining equipment) can be evolved a large distance from the black hole, to null infinity, where the out-state is exactly defined.  By unitarity, the out-state is pure,
so its von Neumann entropy vanishes:
\begin{equation}
S_{b,F-b}=S_F = 0~.
\label{eq-un}
\end{equation}

\subsubsection{The Minable Zone as a Subsystem of the Infalling Vacuum}

The mode $b$ is also a subsystem of the quantum field in a neighborhood ${\cal N}$ of the horizon that includes comparable portions of the interior and the exterior of the black hole. We can choose ${\cal N}$ much larger than the distance of $b$ from the horizon but much smaller than the black hole radius $R$. To be concrete, let $\gamma$ be an infalling geodesic (``Alice'') and let $p,q\in\gamma$ be events at proper time $d$ ($1\ll d\ll R$) before and after the geodesic crosses the black hole horizon. We can define ${\cal N}$ as the causal diamond $I^-(q)\cap I^+(p)$, i.e., the points that can be causally probed by experiments that start after $p$ and end before $q$. I will sometimes refer to the {\em time\/} when Alice falls in, which can be defined as the Schwarzschild time of $q$ to adequate precision. None of these definitions require Alice to fall freely from very far away, just from farther than the mode of interest.

Since no matter has entered the black hole for a time greater than $R\log R$, by the no-hair theorem the region ${\cal N}$ should be free of matter. By the adiabatic theorem, the production of particles localized to ${\cal N}$ is exponentially suppressed in $R/d\gg 1$.  Hence an infalling observer should see the Minkowski vacuum on the scale of ${\cal N}$:\footnote{In fact, the state of ${\cal N}$ is highly entangled with its complement through ultraviolet modes near its boundary. But same type of entanglement would also be present for a causal diamond in exact Minkowski space. Since ${\cal N}$ is much larger than the support of $b$, this entanglement is dominated by modes orthogonal to $b$ and is irrelevant for this discussion.}
\begin{equation}
|\psi \rangle_{\cal N} \approx |0\rangle_M.
\end{equation}

The Minkowski vacuum can be written in Unruh form~\cite{Unr76},
\begin{equation}
|0\rangle_M=C \exp\left(\int d\omega\, e^{-\beta\omega/2} \tilde b_\omega^\dagger b_\omega^\dagger\right) |0\rangle_L |0\rangle_R
\label{eq-unruh}
\end{equation}
where $\beta=2\pi$ and $C$ is a normalization factor. Here $\tilde b_\omega^\dagger$ ($b_\omega^\dagger$) are creation operators for Rindler modes of frequency $\omega$ on the left (right). Transverse momenta have been suppressed for ease of notation. Let us consider a particular right Rindler mode with frequency $\omega$. Evaluating the exponential in Eq.~(\ref{eq-unruh}) for this frequency one finds: 
\begin{equation}
|0\rangle_M\propto |0\rangle_{M-\tilde b_\omega-b_\omega}\otimes \left(\sum_{n=0}^\infty e^{-\beta n\omega/2} |n\rangle_{\tilde b}\otimes |n\rangle_b\right)
\end{equation}
Tracing over the complement of $\mathbf{\tilde b}_\omega \otimes \mathbf{b}_\omega$ trivially gives the pure state in the parentheses: the right mode $b_{\omega}$ is entangled with and purified by the left mode $\tilde b_\omega$. Tracing also over $\mathbf{\tilde b}_\omega$, one obtains a thermal state $\rho_{b_\omega}=(1-e^{-2\pi\omega})\sum e^{-2\pi n\omega} |n\rangle_{b_\omega}\, \mbox{}_{b_\omega}\!\langle n|$. If $\omega\sim O(1)$, this state has von Neumann entropy of order unity:
\begin{equation}
S_b\sim O(1)~.
\label{eq-sb}
\end{equation}

The exact Rindler mode $b_\omega$ considered here has support in the entire right Rindler wedge, and in the Minkowski vacuum, its purification lives only on the left. But consider an approximately stationary wavepacket $b$ of characteristic frequency $\omega$ and size $x\gtrsim \omega^{-1}$, localized strictly in the right wedge at $t=0$. In the state $|0\rangle_M$ the purification of $b$ comes from modes with support both on the left and on the right. 
But the key point remains that in the Minkowski vacuum, the state of $b$ is mixed and cannot be purified without accessing the left wedge. Moreover, by increasing $x$ at fixed $\omega$, the frequency can be made very sharp, so $b\to b_\omega$. It will make no difference below whether the state of $\tilde b b$ in the infalling vacuum is pure or just nearly pure ($S_{\tilde b b}\ll 1$), so for simplicity I will write $S_{\tilde b b}=0$ for wavepackets. Moreover, if the characteristic Rindler frequency of the packet is of order the Unruh temperature, then Eq.~(\ref{eq-sb}) continues to hold at the stated accuracy. To summarize, for the wavepacket $b$ one has
\begin{equation}
S_{\tilde b b}=0~,~~~ S_b\sim O(1)~.
\label{eq-vac}
\end{equation}

Since Minkowski space is a good approximation for the horizon neighborhood ${\cal N}$, and since $b$ is well localized within ${\cal N}$, Eq.~(\ref{eq-vac}) also follows in the case of a wave packet mode $b$ in the near-horizon zone of a black hole. With the appropriate renormalization (unit Killing vector at infinity), the characteristic frequency of the relevant modes is order the Hawking temperature, $R^{-1}$.  The purification $\tilde b$ is the ``partner mode'' in the black hole interior.

With $\tilde b \to X$, $b\to Y$, and $(F-b) \to Z$, we see that unitarity, Eq.~(\ref{eq-un}), and the infalling vacuum, Eq.~(\ref{eq-vac}), correspond to Eqs.~(\ref{eq-sbc}) and (\ref{eq-sabb}), and thus are mutually incompatible. If we insist on the infalling vacuum, then Eq.~(\ref{eq-un}) fails substantially: each Hawking particle carries entropy of order unity, and not a single one can be purified by any other part of the Hawking radiation; so the information about the initial state $|\Psi\rangle$ is lost. Conversely, if we insist on unitarity, then Eq.~(\ref{eq-vac}) fails substantially. The argument applies to any mode $b$ in the near horizon zone, down to a Planck scale cutoff from the horizon. Thus all modes spanning the horizon are in an $O(1)$ excited state at short distances. This is the firewall. 

\subsection{Discussion: the Dual Role of Minability}
\label{sec-dpdisc}

As advertised, this argument has made no assumptions about the age of the black hole, or about the degree of entanglement of the minable zone with any other exterior degrees of freedom. In this respect, it is significantly more general than earlier arguments~\cite{AMPS}. It implies a firewall after the scrambling time, $R\log R$, when the exterior metric has reached its asymptotic form demanded by the no hair theorem, and the apparent horizon nearly coincides with the event horizon that would obtain if no other matter ever enters. If additional matter enters, a new firewall must form a (new) scrambling time later at the new apparent horizon.

The above argument makes more extensive use of the properties of minable modes than~\cite{AMPS}. Here I will aim to address some potential objections and to clarify the role of minability in the firewall argument.

\begin{itemize}

\item Suppose that there existed some yet unknown, fundamental obstruction to mining. Then modes with high angular momentum would almost inevitably fall back into the black hole, because of the angular momentum barrier. But this would not fully resolve the firewall paradox. Modes with low angular momentum escape from the zone on their own account, with probability of order unity. These modes form spherical wavepackets that are sharply localized in the radial direction and close to the horizon, with characteristic size and distance $\lambda\ll R$. Their detection by a local observer is only power-law suppressed~\cite{AMPS} in the detector size over $R$. This alone marks the horizon as a special place. The adiabatic vacuum requires exponential suppression in $R/\lambda$, with no enhancement for wavepackets close to the horizon. 

Because they cannot fully eliminate firewalls, I will not investigate obstructions to mining here; I assume there are none.

\item The minable zone satisfies two important conditions: that the minable modes are a thermally entangled subsystem of the infalling vacuum, and second, that their extraction decreases the entropy of the black hole. If any one of these conditions did not hold, the double purity conflict could not arise. Without the first condition, the entropy of $b$ would not need to be $O(1)$; and to the extent that $b$ did have entropy, its purification would not be concentrated inside the black hole. Without the second condition, one could argue that $b$ only became part of the out-state as a result of actually probing it. I will first discuss their interplay in estabilishing the firewall. Then I will provide two examples that illustrate how at least one of these conditions breaks down in regions other than the Killing horizon of a black hole, where there must not be a firewall.

\item When the minable zone is probed, then both conditions are satisfied. The first condition plays a role in ensuring the second. Because the vacuum is highly entangled, the probability of creating a partner inside the black hole (upon detection of the outside mode) will be of order unity. It can be made arbitrarily close to unity by measuring a wavepacket with sharp Schwarzschild frequency. 

Because of the spacelike character of the Killing vector field behind the horizon, the production of the partner mode decreases the black hole mass, by an amount equal to the mass of the detected particle. Hence, the energy at infinity can be conserved without draining energy from the detector or from the mechanism that holds it in place. The black hole pays for the energy of the mined particle.

\item The first condition makes it important to distinguish the {\em minable\/} zone from the {\em geometric\/} near-horizon zone, defined as all wavepackets with support mainly in the region between $R$ and $3R/2$. The minable modes also have support mainly in this region. But in addition, they have proper wavelength comparable to their proper distance from the horizon, and they do not have much support outside the angular momentum barrier, which will be closer than $3R/2$ for modes with large angular momentum.  

For example, suppose we measure a wave packet of size $\lambda$ localized just inside $3R/2$. By the Reeh-Schlieder theorem, this has some probability of creating a particle behind the black hole horizon. But because such modes have almost no overlap with the minable modes, this probability will be exponentially small. More likely, it will create another particle that is also outside the black hole. The energy for any new particles inside or outside the detector comes from the detector stirring up the vacuum, not from the black hole. This process is trivial and does not conflict with the infalling vacuum; it is like creating any other entangled pair far from the black hole. Hence there is no firewall at $3R/2$, or anywhere far from the black hole. 

\item The second condition is also essential. Consider, for example, the vacuum near the portion of the event horizon inside a collapsing null shell. This spacetime region is exactly flat, and by causality there must not be a firewall. However, an accelerated observer could detect a Rindler mode just outside the event horizon and transport this particle through the shell to future infinity. In the out-state this particle must be purified by the rest of the Hawking radiation, apparently implying a firewall in violation of causality. 

But this process does not mine; it adds energy to the black hole. Mining is possible once the event horizon becomes a Killing horizon, which happens rapidly after collapse or infall but can take up to a scrambling time, $R\log R$. The event horizon inside the shell is not a Killing horizon with respect to the Killing vector field at infinity. %\footnote{However, Eq.~(\ref{eq-outboutb}) does apply as soon as mining is possible. Hence, this should not be faster than the scrambling time. This must be true even for symmetric collapse, where the exact vacuum solution is attained immediately outside the collapsing matter. Otherwise a local observer crossing the event horizon just outside the matter sees a violation of causality: the initial data accessible to the observer need not have produced a black hole, so a firewall on this portion of the horizon would be unacceptable. Naively, a string holding a detector at fixed radius could be prepared early, so as to capture a zone mode immediately outside a symmetrically collapsing dust ball. It will be important to identify the most relevant obstruction. This may be one case where it is relevant that such states have very small entropy compared to the black hole that they form.} 
Hence the partner mode created behind the horizon by the detection~\cite{UnrWal84} can increase the energy of the black hole that forms, compared to what it would have been if no particle had been detected. (Less energy is retained in matter that stays outside the black hole, because of the backreaction that lowers the kinetic energy of the detector.)

As a result there will be more Hawking radiation than if the experiment had not been performed. The detector and exterior environment are entangled with the interior mode, but their purification may reside in the additional Hilbert space. Therefore it is {\em not\/} possible to argue that the exterior mode would have been purified by the smaller Hawking cloud that would have formed if the experiment had not been performed. Hence it is consistent to declare that the region was in the vacuum state prior to the experiment.

The same discussion applies when a black hole grows to larger size. For example, consider a spherical null shell of mass $M-m$ collapsing around a black hole of much smaller mass $m$. The vacuum near the apparent horizon inside the shell can be mined to large distances compared to $m/2$, so the small black hole has a firewall. But the vacuum near the event horizon inside the shell cannot be mined; so a new, larger firewall at $M/2$ need not form until later, on the horizon portion outside of the shell.

\item Returning to the case of successful mining, it is not crucial that the expectation value of the black hole energy decreases by the full energy of the detected particle. In an unclean experiment, the apparatus may add a tiny amount of energy both to the mined particle and to the outside of the black hole. The latter energy may fall into the black hole and partially cancel the energy decrease due to mining. I am merely assuming that mining can be accomplished while keeping these effects small. Then the mined mode cannot be purified by the small excitation that fell into the black hole but {\em can\/} be nearly purified ($S_{b\hat b}\ll 1$) by some scrambled subsystem of the Hawking radiation $\hat b$. Since little energy was added, this subsystem would have been emitted in any case. This implies that $S_{b\hat b}\ll 1$ independently of whether the experiment is actually carried out. Hence, there was already a firewall at the horizon before the experiment.

\item The previous observation is important, because the exterior Rindler or Schwarz\-schild modes formally reach all the way to the geometric event horizon. But semiclassical mining can only reach to the stretched horizon, of order a Planck length outside the geometric horizon, where the local temperature experienced by mining equipment would reach the Planck temperature. However, we can take the wavepackets to have support outside of this region. This will not significantly alter them unless they mainly had support near the stretched horizon to start with. The point is that the characteristic size of the wavepackets depends on the angular momentum but the size of the region that is semiclassically excluded depends on a fixed cutoff. This will also be important in Sec.~\ref{sec-product}.

\item On a final note, the speed-up of the evaporation process that can be achieved with mining~\cite{Bro12} plays no role the firewall argument. What matters is that any minable mode could, in principle, be accessed; and that the energy of the black hole is decreased by its extraction. This implies that its quantum state is determined by the assumed unitarity of the S-matrix. It is irrelevant whether it is actually mined, since in quantum mechanics the state of a system prior to a measurement or coherent manipulation is independent of whether the procedure is carried out or not.

\end{itemize}

\section{Simple Complementarity vs.\ Uniqueness of the Vacuum}
\label{sec-simple}

In this section, I will consider black hole complementarity~\cite{Pre92,SusTho93}, in its simplest form sufficient for evading the xeroxing paradox. I will show that it does {\em not\/} invalidate the argument for firewalls in the previous section. Firewalls remain necessary. This is due to the uniqueness of the infalling vacuum, $|0\rangle_M$. This obstruction was noted in~\cite{Bou12c} (see also~\cite{AveCho12,Bou13,Cho13}); I expand on it here.

\subsection{Quantum Mechanics Argument}
\label{sec-qmsimple}

Again I begin with a general argument at the level of quantum mechanics. In Sec.~\ref{sec-qmdp}, we assumed that $X$ and $Z$ are distinct. Let us now drop this assumption and instead identify $X$ with $Z$ via a unitary map:
\begin{equation}
X \overset{!}{=} Z~;~~~ {\cal H} = X \otimes Y = Z\otimes Y \neq X\otimes Y\otimes Z~.
\end{equation}
This is assumption (a), corresponding to black hole complementarity. If $X$ and $Z$ are the same Hilbert space, then obviously the purity of $XY$ is not only consistent with the purity of $YZ$ but in fact equivalent to it. 

However, suppose that we now demand that (b) $XY$ is in a unique quantum state $|0\rangle$. (This corresponds to the infalling vacuum of a black hole formed from collapse, after a scrambling time.) Finally, we also demand that (c) $YZ$ can be in more than one  distinct quantum state $|\Psi\rangle$. (This would be required by unitarity, since a black hole can be formed from more than one distinct state.) But obviously, (a), (b), and (c) cannot all be true. I will now apply this reasoning to the black hole in more detail.

\subsection{Application to Black Holes}
\label{sec-bhsimple}

Whether or not firewalls form after collapse, unitarity gives rise to the well-known xeroxing problem~\cite{SusTho93b}. The same pure state is apparently present both inside the black hole and in the Hawking radiation, at the same global time. But arguably, the black hole retains accreted information for a scrambling time $R\log R$~\cite{HayPre07}. Then no observer can access both systems simultaneously. So it is consistent to identify the Hilbert space of the matter inside the black hole with the Hilbert space of the Hawking radiation.

In order to address the firewall paradox as exhibited in the previous section, one would need to demand an identification of the interior (assumed to exist) with the Hawking radiation even at times when no matter enters the black hole. This puts too great a burden on the complementarity map, as I will now show.

For any infall time $t$, I demand that an interior exists and that its modes can be identified with a subsystem of $\mathbf{F}$:
\begin{equation}
\mathbf{\tilde b}(t) \overset{!}{=} \mathbf{\hat b} 
\subset \mathbf{F} ~.
\label{eq-id}
\end{equation}
I assume that the map between states in $\mathbf{\tilde b}$ and $\mathbf{\hat b}$ is linear and unitary (i.e., it is invertible and preserves the inner product between pairs of states in $\tilde b$). The unitarity of the map ensures that the unitarity of the local evolution of the collapsing matter inside the black hole is consistent with the unitarity of the S-matrix. (In general, one expects that $\mathbf{\hat b}$ is highly scrambled in $\mathbf{F}$, i.e., a complicated unitary would need to be applied to the physical carriers of the out-state in order to display $\mathbf{\hat b}$ as a physical subsystem.)

There is a new difficulty, however. Unitarity and effective field theory outside the horizon imply that every minable zone mode $b$ involved in the infalling vacuum is part of the out-state Hilbert space $\mathbf{F}$. Hence $\bigotimes'_t \mathbf{B}(t)\subset \mathbf{F}$.\footnote{The prime indicates that this is not a direct product since there will be overlaps between the degrees of freedom present in the zone at different times. For example, for small $t$, $\mathbf{B}(t)\approx \mathbf{B}(t+dt)\approx \mathbf{B}(t)\otimes' \mathbf{B}(t+dt)$. Moreover, unitarity requires zone degrees of freedom to be recycled even over larger time-scales.} Conversely, every Hawking radiation quantum arriving at future null infinity passed through the zone at an earlier time, when it was trivially minable, so $\bigotimes'_t \mathbf{B}(t)\supset \mathbf{F}$. Therefore
\begin{equation}
\mathbf{F} = \bigotimes'_t \mathbf{B}(t)~.
\label{eq-sumb}
\end{equation} 

This result does not mean, nor do I assume, that all of $\mathbf{F}$ is accessible to an exterior observer at a particular single time while the black hole is still present. Similarly, the identification (\ref{eq-id}) only implies that the interior mode $\tilde b(t)$ is part of the out state. It does not mean, nor will I assume, that $\tilde b(t)$ is minable from the zone at any one given time during the evaporation process. 

In the infalling vacuum, the zone mode $b\subset B$ and its interior partner $\tilde b\subset \tilde B$ are in the particular state
\begin{equation}
|0\rangle_{\tilde b b}\propto \sum_{n=0}^\infty e^{-\beta n\omega/2} |n\rangle_{\tilde b}\otimes |n\rangle_b
\label{eq-bb}
\end{equation}
But by Eqs.~(\ref{eq-id}) and (\ref{eq-sumb}), we have $\tilde b b \subset \mathbf{F}$. By Eq.~(\ref{eq-sumb}), the collection of all $b$ modes spans $\mathbf{F}$, so the collection of $\tilde bb$ pairs only introduces redundancy and still spans $\mathbf{F}$.  Mode by mode, Eq.~(\ref{eq-bb}) defines a unique state $|0\rangle_F$ in $\mathbf{F}$. This out-state would be pure, unlike in Hawking's calculation. But it would be independent of the in-state, in violation of unitarity. Information would be lost.

Conversely, if we insist that the out state is a generic state $|\Psi\rangle_F$, then with the identification (\ref{eq-id}), the state of every mode pair $\tilde bb$ will generically be approximately orthogonal to (\ref{eq-bb}). This implies an $O(1)$ deviation from the infalling vacuum. Mode by mode, the probability of seeing no particles differs from unity by an order one quantity. There are $O(A)$ independent minable modes in the semiclassical regime; and this number is dominated by wavepackets with wavelength near the UV cutoff, a distance of order the UV cutoff from the horizon. Hence, the expected number of excited high-frequency modes is $O(A)$, and there is a firewall localized at the horizon.

The vacuum condition, Eq.~(\ref{eq-bb}), will not hold for all zone modes during times when matter enters the black hole. But this does not affect the above argument. Consider a black hole that forms from collapse of a star and then evaporates without further absorption of matter. (The standard picture of a harmless black hole horizon is certainly overthrown if firewalls are present in such black holes.) The above argument begins to apply unchanged at a time of order $R\log R$ after collapse, when the no-hair vacuum configuration should be reached. We may exclude from $\mathbf{F}$ the $\lesssim O(\log R)$ Hawking quanta that will have been emitted during the first scrambling time. These quanta can only carry away $O(\log R)$ qubits of information, but the collapsing star can have up to $O(R^{3/2})$ qubits. The above argument then shows that the remaining Hawking radiation also cannot return the information. 

More generally, any minable mode that is not actually mined or emitted can be excluded from $\bigotimes'_t B(t)$ without invalidating Eq.~(\ref{eq-sumb}). Hence, the modes occupied by infalling matter can be excluded. Then the above argument refers only to modes for which Eq.~(\ref{eq-bb}) can be demanded.  Thus, the out-state becomes highly overdetermined, approximately by the entropy of the infalling matter, if we allow vacuum regions to participate in a unitary complementarity map.

\subsection{Discussion}
\label{sec-simpdisc}

I close this section with two comments.

It is sometimes argued that not all $\exp(S)$ states, $S=A/4$, associated with a black hole of area $A$ can actually be produced. This seems implausible; it would mean that black hole thermodynamics has no standard statistical interpretation, and I argue explicitly against this possibility in Sec.~\ref{sec-product}. But it does not help in any case. Black holes can certainly be formed in many different states. By slowly condensing soft quanta of wavelength comparable to the Schwarzschild radius (``inverse Hawking radiation''), one can produce $\exp(O(A))$ orthogonal states. Rapid collapse of an initially stationary system allows for $\exp(O(A^{3/4}))$. But demanding the infalling vacuum at all times leads to a unique out-state. One could ``reserve'' a fraction $1-x$ (with $x\sim O(1)$ or $x\sim A^{-1/4}$) of the degrees of freedom in $F$ for factors of the form Eq.~(\ref{eq-bb}). But all of the Hawking radiation passes through the zone. Even a small fraction $A^{-1/4}$ of s-waves that are firewalls while in the zone constitutes an unacceptable violation of the equivalence principle, since their characteristic size will be much smaller than $R$ near the horizon. This is a horizon marker, in violation of the equivalence principle. Deviations from the adiabatic vacuum in quanta of wavelength much less than than the curvature radius must be exponentially suppressed.

It is important to note that complementarity, understood as a unitary map relating the interior to the out-state, is perfectly consistent, {\em if we do not insist on the infalling vacuum} except in those places where the existence of a firewall is excluded by causality. This weaker requirement will not overdetermine the out-state. When a system collapses to form a black hole, it enters the interior before the firewall forms; this region must have an image in $\mathbf{F}$, by unitarity of the S-matrix. Similarly, if the black hole grows due to accretion, matter enters the interior of the new, larger black hole before hitting the (old) firewall. (Later a new firewall forms at the larger horizon.) The accreted information contains, by unitarity, is also in $\mathbf{F}$ and so there is a ``pull-back-push-forward procedure''~\cite{FreSus04,BouSus11,Sus12c} that establishes a unitary complementarity map between interior and Hawking radiation. The vacuum (i.e., the fact that certain modes were not excited in the infalling matter) can be included in this map, while the pull-back-push forward procedure remains well-defined. However, the procedure becomes ill-defined a scrambling time~\cite{Sus12c} after accretion, because the backward evolution out of the black hole would involve transplanckian frequencies. The region between the old firewall and the new event horizon ceases to be accessible, so pull-back-push-forward based on the most recent infall does not constrain the state on this part of the horizon, and a unitary complementarity map need not include this region. This is consistent with a new firewall having formed after the scrambling time.

\section{Strong Complementarity vs. Linearity}
\label{sec-strong}

In this section, I argue that simple complementarity (i.e., ``pull-back-push-forward'', as described in the previous paragraph), cannot be generalized or extended so as to eliminate firewalls. It would seem problematic to relax the unitarity of the complementarity map, since it is then not clear how the map can remain consistent with the unitarity of the S-matrix. And the previous section showed that the map cannot be extended to include the vacuum behind the horizon more than a scrambling time after the most recent accretion. But I will not use either of these arguments here. Instead, I will present evidence that  the infalling vacuum cannot be recovered in any case, at the full level of generality required by the equivalence principle, no matter how the map is defined. (As discussed in the introduction, my viewpoint is binary. If the equivalence principle is not fully recovered, then it remains violated. I regard this as the only relevant criterion.)

The most general map one can consider is a nonunitary map from $\mathbf{F}$ to $\mathbf{\tilde B}(t) \mathbf{B}(t)$ that takes $|\Psi \rangle_F  \to |0\rangle_{\tilde B B}$, for all $|\Psi\rangle_F$.\footnote{The map need not involve all of $\mathbf{F}$, and it could depend on $t$.  Observers whose infall time differs by more than the scrambling time cannot compare their experiences at the horizon. Hence one could take the viewpoint that the equivalence principle is recovered as long as a map can be found for any one observer, such the horizon is in the vacuum when and where {\em they} cross it. Here I grant this flexibility, which is called {\em observer complementarity}~\cite{Bou12cV1} or {\em strong complementarity}~\cite{HarHay13}. I argue that the approach falls short in any case.} 
However, both the total Hawking radiation $F$, and the neighborhood of the horizon $\tilde B B$ contain the minable zone $B$ as subsystems. In general, such a map would assign two inconsistent states to $B$; for example, if $|\Psi \rangle$ factorizes, the state of the zone is pure; yet the state of the zone in the vacuum $|0\rangle_M$ must be mixed. 

If $|\Psi\rangle_F$ is random with respect to the Haar measure, then with overwhelming probability the state of $B$ alone is nearly exactly thermal. I will begin by discussing maps that attempt to exploit this fact, in Sec.~\ref{sec-donkey}. I will introduce a toy model in Sec.~\ref{sec-twobit} to illustrate the issues explicitly. In Sec.~\ref{sec-product} I argue that while nonthermal states of $B$ are Haar-rare they are not Boltzmann-rare; they span the full microcanonical ensemble. In Sec.~\ref{sec-strongdisc} I argue that this fact alone constitutes a violation of the equivalence principle, independently of the status of thermally entangled states.

\subsection{Donkey Map}
\label{sec-donkey}

There do exist states $|\Psi\rangle_F\in \mathbf{F}$, such that the state of $B$, regarded as a subsystem of $F$, is the same as the thermal state of $B$ as a subsystem of the vacuum:
\begin{equation}
\rho_B=\mathrm{tr}_{F-B} |\Psi\rangle_F\, \mbox{}_{F} \langle \Psi | = \mathrm{tr}_{\tilde B} |0\rangle_{\tilde B B}\, \mbox{}_{\tilde B B} \langle 0|~.
\label{eq-coincide}
\end{equation}
In this case a many-to-one map $|\Psi\rangle_F \to |0\rangle_{\tilde B B}$ can be realized as a map just between $\tilde B(t)$ and $\hat B(t)$, the purification of $B$ in $F$:
\begin{equation}
\bm{D}(|\Psi \rangle; t): \mathbf{\hat B}(t)\to \mathbf{\tilde B}(t)~.
\label{eq-dmap}
\end{equation}

To distinguish it from the state-independent and infall-time-independent unitary map of traditional complementarity, Eq.~(\ref{eq-id}), I will call this construction a {\em donkey map}. We will see explicitly in the example below how $\bm{D}$ must depend on the full out-state $|\Psi \rangle$. Viewed as a map from $\mathbf{F}$ to $\mathbf{B\tilde B}$, the donkey map is many-to-one,\footnote{This was first criticized in Ref.~\cite{Bou12c} (see also~\cite{AveCho12,Bou13}). Following, e.g., \cite{PapRaj12,VerVer12,Sus13,NomVar13,VerVer13a,VerVer13b,MalSus13}, some difficulties associated with state-dependence were further elaborated in Ref.~\cite{AMPSS,MarPol13}. The present nomenclature is inspired by recent discussions~\cite{VerTalk,MarPol13}. What I here call a donkey map does not fully represent the content of, nor differentiates between, Refs.~\cite{PapRaj12,VerVer12,Sus13,NomVar13,VerVer13a,VerVer13b,MalSus13}. But I argue that none of these proposals can eliminate firewalls in product states.}  since it always results in the infalling vacuum independently of the out-state. 

Eq.~(\ref{eq-coincide}) is indeed satisfied to high accuracy for Haar-typical states $|\Psi\rangle_F$~\cite{Pag93}. Hence, extant arguments have mainly focussed on this case. For example, Ref.~\cite{AMPS} noted that for old enough black holes both systems, $\tilde B$ and $\hat B$, are semiclassically accessible to the infalling observer; hence it is inconsistent to identify them in any manner. (By contrast, in the conventional complementarity of Sec.~\ref{sec-simple}, the two systems that are identified are not semiclassically accessible to any one observer.) The counterargument that $\hat B$ may not be computationally accessible~\cite{HarHay13} has been questioned in Ref.~\cite{AMPSS}. The issue remains controversial~\cite{VerVer13a,VerVer13b,MalSus13,MarPol13,Bou13b}.

Here I will argue that strong complementarity falls short in any case, for a different reason~\cite{Bou13}. I focus on the problem noted earlier: $B(t)$ can only have one state, so it must be invariant under the map from $\mathbf{F}$ to $\mathbf{B}(t)\mathbf{\tilde B}(t)$. Hence the map can only connect $\tilde B(t)$ and $F-B(t)$ nontrivially.  But if the state of $B(t)$ is pure, then by Eq.~(\ref{eq-ss}), no identification of $\mathbf{\tilde B}(t)$ with any subspace of $\mathbf{F}\oslash\mathbf{B}(t)$ can achieve the vacuum. Pure states in $\mathbf{B}(t)$ form a complete basis of $\mathbf{F}$ and hence are not Boltzmann suppressed, as acceptable deviations from the adiabatic vacuum must be. Thus, the equivalence principle is violated: the horizon is a special place. 

\subsection{Two Qubit Example}
\label{sec-twobit}

A simple example will illustrate the donkey map, its structure, and its shortcomings. I model the zone, $B(t)$, as a single qubit $b$. The infalling vacuum is modeled as the EPR state
\begin{equation}
|0\rangle_{b\tilde b} = |0\rangle_b |0\rangle_{\tilde b}+ |1\rangle_b |1\rangle_{\tilde b}~.
\label{eq-epr} 
\end{equation}

\paragraph{State-dependence of the map} I begin by considering the case where $|\Psi \rangle$ is maximally entangled. For simplicity, I take $\mathbf{F}=\mathbf{b}(t) \mathbf{\hat b}(t)$. (In general $\mathbf{F}$ may contain additional Hilbert space factors that do not participate in the map or whose participation itself depends on the state $|\Psi \rangle$.) This toy model and state could represent a young black hole (negligible radiation has been emitted), if we take $\mathbf{\hat b}$ to be the semiclassically inaccessible Planckian modes near the horizon (assumed to contain precisely half of the degrees of freedom, for simplicity). Or it could be interpreted as a half-evaporated black hole: $\mathbf{\hat b}$ would represent the Hilbert space of the early Hawking radiation. In this case, we would assume that all of the zone is semiclassically accessible, for simplicity. 

Each of the four Bell states that span $\mathbf{F}$ can be converted to $|0\rangle_M$ by a map that acts only on $\tilde b$. But as the reader can easily verify, for each such state of $\mathbf{F}$, the map from $\tilde b$ to $\hat b$ must be chosen a different Pauli matrix:
\begin{equation} 
\begin{array}{ c | c | c }
\sqrt{2}\, |\Psi \rangle_{b\hat b} & \bm{D}(|\Psi\rangle): \mathbf{\hat b} \to \mathbf{\tilde b} & \sqrt{2}\, |0\rangle_{b\tilde b}\\
\hline\hline
\begin{array}{c} \mbox{} \\  |0\rangle_b |0\rangle_{\hat b}+ |1\rangle_b |1\rangle_{\hat b} \end{array} & 
\begin{array}{c}\mathbf{1} \\ \longrightarrow \end{array} & 
\begin{array}{c} \mbox{} \\ |0\rangle_b |0\rangle_{\tilde b}+ |1\rangle_b |1\rangle_{\tilde b} \end{array}\\ \hline
\begin{array}{c} \mbox{} \\  |0\rangle_b |0\rangle_{\hat b}- |1\rangle_b |1\rangle_{\hat b} \end{array} & 
\begin{array}{c}\bm{\sigma_z} \\ \longrightarrow \end{array} & 
\begin{array}{c} \mbox{} \\ |0\rangle_b |0\rangle_{\tilde b}+ |1\rangle_b |1\rangle_{\tilde b} \end{array}\\ \hline
\begin{array}{c} \mbox{} \\  |0\rangle_b |1\rangle_{\hat b}+ |1\rangle_b |0\rangle_{\hat b} \end{array} & 
\begin{array}{c}\bm{\sigma_x} \\ \longrightarrow \end{array} & 
\begin{array}{c} \mbox{} \\ |0\rangle_b |0\rangle_{\tilde b}+ |1\rangle_b |1\rangle_{\tilde b} \end{array}\\ \hline
\begin{array}{c} \mbox{} \\  |0\rangle_b |1\rangle_{\hat b}- |1\rangle_b |0\rangle_{\hat b} \end{array} & 
\begin{array}{c} i\bm{\sigma_y} \\ \longrightarrow \end{array} & 
\begin{array}{c} \mbox{} \\ |0\rangle_b |0\rangle_{\tilde b}+ |1\rangle_b |1\rangle_{\tilde b} \end{array}\\
\end{array}
\label{eq-donkey}
\end{equation}
Note that in every state $|\Psi \rangle_{b\hat b}$ considered above, the accessible mode $b$ is in the same state that it has in the vacuum, 
\begin{equation}
\rho_b=\frac{1}{2}(|0\rangle_b\, \mbox{}_{b} \langle 0| + |1\rangle_b\, \mbox{}_{b}\langle 1|)~,
\end{equation}
as required.

\paragraph{Product states} The Bell states in Eq.~(\ref{eq-donkey}) form a complete basis of $\mathbf{F}$, but so do product states. In this case $b$ is already pure by itself, so the Hilbert space $\mathbf{\hat b}$ is empty, and the donkey map cannot defined by Eq.~(\ref{eq-dmap}). Moreover, any map from $\mathbf{F}$ to $\mathbf{b} \mathbf{\tilde b}$ must leave $b$ invariant. It can only act on the complement of $b$ in $F$, denoted $F-b$, with Hilbert space factor $\mathbf{F} \oslash \mathbf{b}$. Hence, the product structure is preserved, and the vacuum cannot be obtained with any choice:
\begin{equation}
\begin{array}{ c | c | c }
|\Psi \rangle_F & \bm{D}(|\Psi\rangle): \mathbf{F}\oslash \mathbf{b} \to \mathbf{\tilde b} & |0\rangle_{b\tilde b}\\
\hline\hline
\begin{array}{c} \mbox{} \\  |0\rangle_b |0\rangle_{F-b} \end{array} & 
\begin{array}{c}\rm{any} \\ \longrightarrow \end{array} & 
\begin{array}{c} \mbox{} \\ |0\rangle_b |\alpha\rangle_{\tilde b} \end{array}\\ \hline
\begin{array}{c} \mbox{} \\  |0\rangle_b |1\rangle_{F-b} \end{array} & 
\begin{array}{c}\rm{any} \\ \longrightarrow \end{array} & 
\begin{array}{c} \mbox{} \\ |0\rangle_b |\beta\rangle_{\tilde b} \end{array}\\ \hline
\begin{array}{c} \mbox{} \\ |1\rangle_b |0\rangle_{F-b} \end{array} & 
\begin{array}{c}\rm{any} \\ \longrightarrow \end{array} & 
\begin{array}{c} \mbox{} \\ |1\rangle_b |\gamma\rangle_{\tilde b} \end{array}\\ \hline
\begin{array}{c} \mbox{} \\  |1\rangle_b |1\rangle_{F-b} \end{array} & 
\begin{array}{c} \rm{any} \\ \longrightarrow \end{array} & 
\begin{array}{c} \mbox{} \\ |1\rangle_b |\delta\rangle_{\tilde b} \end{array}\\
\end{array}
\label{eq-puredonkey}
\end{equation}
where the states $|\alpha\rangle_{\tilde b},\ldots$ depend on the arbitrary map $\bm{D}$.

It is worth restating this point. In every state $|\Psi \rangle_F$ in the above basis, the accessible mode $b$ is in a pure state (either $|0\rangle_b$ or $|1\rangle_b$, though of course a different choice could have been made). This state must be preserved: because $b$ can be measured before crossing the horizon, all observers must agree on its state. Indeed, the map $\bm{D}$ acts only on $\mathbf{F}\oslash \mathbf{b}$. But this means that the neighborhood of the vacuum is also in a product state. The overlap of any product state with the vacuum state, Eq.~(\ref{eq-epr}), differs from unity by a term of order unity. Thus the probability for encountering a particle at the horizon is substantial: a firewall.

\subsection{Product States Form a Complete Basis}
\label{sec-product}

In this subsection I argue that Eq.~(\ref{eq-puredonkey}) of the toy model correctly captures a property of the Hilbert space of a black hole: There exists a complete (highly non-unique) basis, such that every basis element is a product state of the zone $B$ and any other degrees of freedom that might be associated with the black hole. Because $B$ by itself is already pure, every such state has a firewall. A map that identifies the exterior purification of $B$ with the interior $\tilde B$ cannot help since it has no space to act on. 

\subsubsection{Black Hole Thermodynamics has a Statistical Interpretation}

I will assume that black hole thermodynamics~\cite{Bek72,BarCar73,Haw75} is valid.\footnote{This might be questioned if there are firewalls. But the argument for firewalls is by contradiction and so assumes their absence; then black hole thermodynamics stands on a solid footing.} In particular, black holes satisfy a first law, Eq.~(\ref{eq-fl}), with (for Schwarzschild)
\begin{equation}
S=\frac{A}{4}~,~~~ T=\frac{1}{4\pi R}~,~~~E=\frac{R}{2}~.
\end{equation}

Moreover, I assume that black hole thermodynamics has a statistical interpretation, as unitarity demands.\footnote{Note that this assumption refers only to the black hole as probed by an exterior observer. It does not prejudice what other system the interior modes might be identified with. Thus, I do not assume the ``proximity postulate''~\cite{Sus13}.} To be concrete, I assume that black holes share the following standard properties of the canonical and microcanonical ensembles of ordinary thermodynamic objects, such as the air in a sealed room, or a cavity filled with electromagnetic radiation: 

\paragraph{Canonical Ensemble} The entropy as a function of energy, $S(\langle E \rangle )$, can be macroscopically determined by controlling the temperature, measuring $\langle E \rangle (T)$, and integrating
\begin{equation}
\frac{dS}{d \langle E \rangle}=\frac{1}{T}~.
\label{eq-fl}
\end{equation}
The microscopic interpretation is
\begin{equation}
S=-\mathrm{tr}\,\rho\log\rho~,
\label{eq-sr}
\end{equation}
where 
\begin{equation}
\rho=\sum_i p_i |i\rangle \langle i|~.
\label{eq-rhodef}
\end{equation}
Here $p_i\propto \exp(-E_i/T)$ is the probability of finding the system in the state $|i\rangle$ with energy $E_i$. The Hilbert space ${\cal H}$ in which $\rho$ acts is generally of infinite dimension (e.g., a Fock space). The energy expectation value is
\begin{equation}
\langle E \rangle=\mathrm{tr}_{\cal H} \, \rho E
\end{equation}

\paragraph{Microcanonical Ensemble} The microcanonical ensemble consists of states with energy in a narrow range around $\langle E \rangle$. Its density matrix is proportional to the unit matrix:% in the Hilbert space $\bar{\cal H}$ spanned by these states:
\begin{equation}
\bar\rho = {\cal N}^{-1}\sum_{E_i\approx E} |i\rangle \langle i| 
\end{equation} 
where ${\cal N}$ is the finite dimension of the Hilbert space $\bar{\cal H}$ spanned by these states.  

\paragraph{Thermodynamic Limit} In the thermodynamic limit, the entropy of the canonical ensemble is dominated by states of energy $\langle E \rangle$, and so agrees with the microcanonical entropy
\begin{equation}
\bar S \equiv \log {\cal N} = S~.
\label{eq-sn}
\end{equation}

The statistical interpretation implies that canonical and microcanonical ensembles with these properties exist for a black hole. This requires suitable boundary conditions, such as a box~\cite{Haw76}. Anti-de Sitter space makes for a good box~\cite{HawPag83}, but any small enough box will do. The box is needed only to prepare the black hole; the timescale for this may be very long. A firewall for a black hole in a box seems no more acceptable than in any other setting; the equivalence principle is violated eiter way. Moreover, the box can be removed just before infall, after an appropriate state is prepared, since the properties of the zone cannot change substantially on the timescale $R$. In order to keep the discussion general, I will not appeal to a CFT dual~\cite{Mal97}, nor to any other assumptions about the fundamental nature of the microscopic degrees of freedom.

%The canonical ensemble is defined by a mixed density operator $\hat\rho \propto e^{-\hat E/T}$ acting in a Hilbert space ${\cal H}$ of infinite dimension, with von Neumann entropy $S=A/4$. The microcanonical ensemble consists of states of approximately fixed energy; it spans a Hilbert space $\bar {\cal H}$ of finite dimension $e^{\bar S}$, with $\bar S=S$ in the thermodynamic limit (large black holes). 

\subsubsection{The Minable Zone is a Subsystem}

An important question is what constitutes the black hole, i.e., what physical degrees of freedom correspond to its Hilbert space. For an ordinary thermodynamic system, one only considers microstates consistent with the macroscopic (``coarse-graining'') conditions imposed (for example, excitations confined to a cavity of some radius). For the black hole, I take these conditions to be those assumed in the derivation of the laws of black hole mechanics: the metric is a vacuum Schwarzschild (or Kerr) solution with energy $E$ as measured by a distant observer. 

The energy is, in a sense, purely a feature of the geometry; the stress tensor vanishes everywhere. Yet, black holes share a key property with other thermodynamic systems: the energy $E$ can be lowered by mining, consistent with the First Law. Hence, the minable modes in the zone, $B$, {\em must\/} be considered a subsystem of the black hole. 

On the other hand, the black hole interior, $\tilde B$, cannot be directly accessed by an external observer. But these are the only observers to whom the horizon entropy, temperature, and the mass of the black hole have operational meaning. Hence it does not seem natural to include interior field theory modes in ${\cal H}$. 

However, I will be considering ideas in which the interior is identified with distant degrees of freedom accessible to an exterior observer. My goal is to argue against these proposals, so it is important that they are not trivially excluded from the start. In any case, it is not obvious that $B$ alone constitutes the black hole's degrees of freedom. Therefore I will allow ${\cal H}$ to contain a ``hidden'' factor $\mathbf{H}$, some or all of which may be associated with the interior.  

Without loss of generality, the black hole Hilbert space in which the thermal density matrix $\rho$ acts can thus be written as
\begin{equation}
{\cal H} = \mathbf{H}\otimes \mathbf{B}~. 
\label{eq-prod}
\end{equation}

\subsubsection{The Canonical Zone Entropy is Additive}

Consider first the canonical ensemble. In a field theory calculation without a cutoff, the thermal entropy of the zone, $S_B$, would diverge due to ultraviolet modes near the horizon. But black hole thermodynamics dictates that the total thermal entropy is $S=A/4$. This is consistent with a Planck scale cutoff in the field theory. However, there are a number of possibilities for how the zone contributes to the total entropy $S$.

The simplest possibility, which I will {\em not\/} assume, is that the Planck cutoff, when properly derived from the underlying theory, is such that the zone is entirely responsible for the Bekenstein-Hawking entropy: $S_B=S$. Then all states in the microcanonical ensemble would trivially be pure states of the zone (and so have a firewall). This may be the case: in standard gravity, effective field theory should be a good approximation up to the Planck scale, so the canonical entropy of the zone alone satisfies $S_B\sim O(A)$. 

However, suppose we adopt a conservative definition of ``semiclassical'', a relatively low enough frequency cutoff, e.g. $\omega_c= 1/100$ in Planck units. Then the semiclassically minable zone $B$ would represent only a subsystem of the black hole. 
Hence, the black hole may contain additional degrees of freedom $H$, which are hidden from the outside observer or protected from semiclassical access. This is the most general possibility. Subadditivity of the entropy of subsystems implies that
\begin{equation}
S_B+S_H\geq S~.
\label{eq-mut}
\end{equation}

In fact, however, the canonical entropies of the zone and the hidden degrees of freedom must be additive: 
\begin{equation}
S_B+S_H= S~.
\label{eq-nomut}
\end{equation}
This follows from general properties of the canonical ensemble, and from the definition of $B$ as consisting of degrees of freedom that can be extracted from the black hole. Consider an ordinary system, such as a cavity filled with blackbody radiation, in a thermal state with finite entropy $S$. Let ``$B$'' and ``$H$'' be the left and right half of the cavity. Defined strictly geometrically, the von Neumann entropy of ``$B$'' diverges due to the entanglement entropy of ultraviolet modes near the dividing surface. However, an entangled mode cannot escape from the cavity even if a hole is opened or a wire inserted. If it did, then in equilibrium it would be replaced by energy from the heat bath. But the replacement would not have the same entanglement with ``H''. As a result, the total entropy of the cavity would change, at fixed temperature. In this sense, entangled systems inside a thermal system cannot participate in the canonical ensemble. By contrast, if the division of the cavity is implemented by inserting a wall with appropriate boundary conditions, the entanglement will be eliminated. Then the entropy will satisfy $S_B+S_H=S$, and all parts of ``$B$'' can be exchanged with the heat bath. 

But the latter case is the one analogous to the black hole, because $B$ was {\em defined\/} to consist of minable modes. If the inequality (\ref{eq-mut}) was strict, then $B$ and $H$ would have mutual information. Then the exchange of $B$ with heat bath degrees of freedom would lead to a build-up of mutual information between the heat bath and the black hole, and to an increase in the canonical entropy of the black hole. This is impossible.

In the black hole case, it is important that $B$ and $H$ are allowed to interact, because only $B$ can be semiclassically coupled to a heat bath. If they could not interact, then the information in $H$ could not get out in the evaporation process. This interaction presents no obstruction, however, because they can only interact at the stretched horizon. That is, information from $H$ must enter $B$ through modes that get stretched below the cutoff, and vice-versa. Otherwise, $H$ would not be hidden, which would simplify my argument.  As discussed in Sec.~\ref{sec-dpdisc}, the minable wavepackets can be taken to have support only away from the stretched horizon without affecting their crucial properties. They will still carry nearly as much energy as a Schwarzschild frequency eigenstate, and they will have $O(1)$ entropy in the infalling vacuum.

In fact the difficulty introduced by this interaction is no greater than in any ordinary thermodynamic system. For example, instead of the divided cavity, consider now a cavity with a ball ``$H$'' in the center. The ball is coupled to the heat bath only through the radiation ``$B$'', which has only short range interactions with ``$H$'', much shorter than the distance between the ball and the outer wall of the cavity.  The canonical ensemble still factorizes if we slightly redefine ``$B$'' to exclude the interaction region.

\subsubsection{The Microcanonical Ensemble Factorizes}

Additivity of the canonical entropy implies that $B$ and $H$ can be treated as independent systems coupled to the same heatbath. Therefore, even if there exist hidden degrees of freedom $H$, one can go over to the microcanonical ensembles for $B$ and $H$ independently. The microcanonical ensemble for $B$ spans a Hilbert space $\mathbf{\bar B}$ of finite dimension $\exp(S_B)$, where $S_B\sim O(A)\leq S$ is the canonical entropy of $B$. If $S_B<S$, then there also exists a microcanonical Hilbert space for the hidden degrees of freedom, $\mathbf{\bar H}$, with dimension $\exp(S_H)$, where $S_H=S-S_B$. The full microcanonical ensemble for the black hole is $\mathbf{\bar B}\otimes \mathbf{\bar H}$. Like any outer product, it admits a basis consisting entirely of product states:
\begin{equation}
\{|i\rangle_{\bar H}\otimes |j\rangle_{\bar B}\}~,
\end{equation}

In each basis element, the state of the zone is pure. Thus, if an interior exists, the zone $B$ and interior $\tilde B$ together are in a product state
\begin{equation}
|\alpha\rangle_{\tilde B}\otimes |j\rangle_{\bar B}
\label{eq-bbprod}
\end{equation}
i.e., completely unentangled. This is true whether the interior is modelled as independent degrees of freedom, or identified with degrees of freedom in $H$ or in the distant Hawking radiation, by any map, unitary or not.

Of course, it is not important that the state of $B$ be exactly pure. For a firewall, it suffices that its entropy is much smaller than its canonical entropy, in each basis element. As discussed above, the division into subsystems $B$ and $H$ could be somewhat blurred by boundary effects. This is true for any thermodynamic system, as in the example of the radiation-filled cavity with a ball ``$H$'' in the center. But if both subsystems are large and interactions are localized to their shared boundary, then independently of the detailed definition of $B$, one can find a complete basis of the microcanonical ensemble of $BH$ such that in every state, the entropy of $B$ is much smaller than it would be in the canonical ensemble. 

One can easily estimate the probability of seeing no particles in a general product state. For each mode, the infalling vacuum is the entangled pure state
\begin{equation}
\sum_{n=0}^\infty e^{-\beta n\omega/2} |n\rangle_{\tilde b}\otimes |n\rangle_b\
\end{equation}
In each basis element identified above, the states of $b$ and $\tilde b$ form a product state, which may be mixed or pure. Its overlap with the vacuum is maximized if this product state is $|0\rangle_b |0\rangle_{\tilde b}$, when the probability of observing the vacuum is $1-e^{-2\pi\omega}$. Hence the probability of observing a particle is at least $e^{-2\pi\omega}$, mode by mode, or about $e^{-1}$ for modes with thermal wavelength. The minable zone contains of order $A$ independent modes (with a Planck scale cutoff). This is a firewall; the probability of seeing no particles at all (the infalling vacuum) is $\exp[-O(A)]$.\footnote{Of course, the product state obstruction also applies in the more restrictive setting of a linear unitary state-independent map. In this case, I have already exhibited a different obstruction (Sec.~\ref{sec-simple}) which applies to arbitrary states. } 

\subsection{A Complete Basis of Firewall States Violates the Equivalence Principle}
\label{sec-strongdisc}

Product states span the whole Hilbert space, but they do not constitute it, since the product form is not preserved by linear combinations. This appears to leave a loophole: for a large system, product states are extremely rare with respect to the Haar measure~\cite{Pag93}. Generic (Haar-random) pure states are highly entangled. In such states, the state of $B$ will be a thermal density matrix, identical to the state one would obtain from the infalling vacuum. This is precisely the setting where a donkey map can restore the infalling vacuum even though $B$ is purified into a (naively) different state by a (naively) different system.\footnote{In a separate publication~\cite{Bou13b}, I argue that for generic states, the donkey map is overkill: the infalling state cannot deviate from the vacuum even in situations where the equivalence principle demands that it must.}
But the mere existence of a complete basis of firewall states, which has been demonstrated in the previous subsection, is incompatible with the properties expected of deviations from the vacuum in curved spacetime regions.

Consider a freely falling detector in a vacuum spacetime region with curvature radius $R$. The adiabatic theorem dictates~\cite{BirDav} that the probability for detecting particles of characteristic frequency $\omega\gg R^{-1}$ is suppressed exponentially, like $e^{-O(\omega R)}$. This is satisfied by black holes in the infalling vacuum: the zone is at a temperature $T\sim R^{-1}$, so Boltzmann suppression of energetic quanta yields a result consistent with the adiabatic theory. In particular, the atypical states where the black hole emits a high-energy object are {\em not\/} Haar-suppressed. Unlike product states, they do constitute a subspace; and this subspace has dimension exponentially smaller than $e^{O(A)}$. These are the atypical states consistent with the adiabatic vacuum. By contrast, all product states are firewall states, but they do not constitute a Hilbert space. Yet, they span the entire Hilbert space ${\cal H}$ and so are not Boltzmann suppressed.

Moreover, particle detection in the adiabatic vacuum in curved space is not sharply localized. A soft particle of energy $R^{-1}$ might be encountered anywhere in a spacetime region with curvature radius $R$. A freely falling observer may also find more energetic excitations but they must be exponentially suppressed. Such particles are indeed predicted with the appropriate Boltzmann suppression in the Hawking temperature, and again they can occur anywhere in the zone. By contrast, the firewall in the product state basis is sharply localized to the apparent horizon of the black hole. The location is the same for all $\exp(A/4)$ firewall states. There is no comparably sharp effect anywhere else, for any other distinct choice of Haar-rare states. At the level of the theory, this violates the equivalence principle: the horizon is a special place. Then it is not clear what is gained by arguing that entangled states are smooth. 

\acknowledgments I would like to thank the organizers of the CERN workshop on Black Hole Horizons and Quantum Information, March 2013, where the main results of this paper were first presented. I have benefitted from valuable discussions with many colleagues, especially S.~Giddings, D.~Harlow, D.~Mainemer~Katz, D.~Marolf, J.~Polchinski, V.~Rosenhaus, D.~Stanford, and L.~Susskind. This work was supported by the Berkeley Center for Theoretical Physics, by the National Science Foundation (award numbers 1002399, 0855653 and 0756174), by fqxi grant RFP3-1004, by ``New Frontiers in Astronomy and Cosmology'', and by the U.S.\ Department of Energy under Contract DE-AC02-05CH11231.

\bibliographystyle{utcaps}
\bibliography{all}

\end{document}